\documentstyle[aps,prl,multicol,epsfig]{revtex}

\begin{document}

\title{Signatures of Prelocalized States in Classically Chaotic Systems}
\author{A. Ossipov, Tsampikos Kottos, and T. Geisel
\\
Max-Planck-Institut f\"ur Str\"omungsforschung und Fakult\"at Physik
der Universit\"at G\"ottingen,\\
Bunsenstra\ss e 10, D-37073 G\"ottingen, Germany}
\maketitle

\begin{abstract}
We investigate the statistics of eigenfunction intensities ${\cal P}(|\psi|^2)$ in 
dynamical systems with classical chaotic diffusion. Our results contradict some
recent theoretical considerations which challenge the applicability of field theoretical 
predictions, derived in a different framework for diffusive disordered samples. For 
two-dimensional systems, the tails of ${\cal P}(|\psi|^2)$ contradict the results 
of the optimal fluctuation method, but agree very well with the predictions of the 
non-linear $\sigma-$model.
\\

\end{abstract}

\begin{multicols}{2}

The statistical properties of wavefunction intensities have sparked a great deal of 
research activity in recent years. These studies are not only relevant for mesoscopic 
physics \cite{AKL91,MK95,FM94,FE95,M97,M00,SA97,MKW90,UMRS00,N01}, but also for 
understanding phenomena in areas of physics, ranging from nuclear \cite{ZBFH96} 
and atomic \cite{BUF96} to microwave physics \cite{SS90} and optics \cite{NS97}. 
Experimentally, using microwave cavity technics it is possible to probe the 
microscopic structure of electromagnetic wave amplitudes in chaotic or disordered 
cavities \cite{SS90}. Recently, the interest in this problem was renewed when new
effective field theoretic techniques were developed for the study of the distribution 
of eigenfunction intensities ${\cal P} (|\psi|^2)$ of {\it random} Hamiltonians. As 
the disorder increases, these results predict that, the eigenfunctions become 
increasingly non-uniform, leading to an enhanced probability of finding anomalously 
large eigenfunction intensities in comparison with the random matrix theory (RMT) 
prediction. Thus, the notion of {\it prelocalized} states has been introduced 
\cite{AKL91,MK95,FM94,FE95} to explain the appearance of long tails in the 
distributions of the conductance and other physical observables \cite{AKL91}.

Up to now all theoretical predictions \cite{AKL91,MK95,FM94,FE95,M97,M00,SA97} and
numerical calculations \cite{UMRS00,N01} apply to disordered systems and are based 
on an ensemble averaging over disorder realizations. Their validity, however, for a 
quantum {\it dynamical} system (with a well defined classical limit) that behaves 
diffusively is not evident. Furthermore, based on an argument put forward in \cite{M00} 
(see also \cite{AZ96}), the far tail of ${\cal P}(|\psi|^2)$ is due to rare realizations 
of the disorder potential, and therefore requires an exponentially large number of 
eigenfunctions, which can only be accounted by disorder averaging. Here instead we study 
the statistical properties of eigenfunctions in a {\it dynamical} model without 
introducing any ensemble averaging.  Our main conclusion is that in a generic dynamical 
system with classical diffusion, ${\cal P}(|\psi|^2)$ is described quite well by the 
nonlinear $\sigma-$model (NLSM). We point out here that between the various theoretical 
works there is a considerable disagreement about the parameters that control the 
shape of ${\cal P}(|\psi|^2)$ and their dependence on time-reversal symmetry (TRS). 
More specifically, the NLSM suggests that the tail of  ${\cal P}(|\psi|^2)$ in two
dimensions ($2d$) is sensitive to TRS \cite{FE95,M97,M00}, while a direct optimal 
fluctuation (DOF) method predicts a symmetry independent result \cite{SA97}. Recent 
numerical calculations \cite{UMRS00} on the Anderson model seem to support the 
latter theory. This controversy, was an additional motivation for the present work.

%==============================================================================================

In the present article, we numerically study the distribution of intensities of the Floquet-
states of the kicked rotor (KR) on a torus \cite{I90} and its $2d$ generalization
\cite{DF88}. Our system is defined by the time-dependent Hamiltonian 
\begin{eqnarray}
\label{ham}
H=H_0 + k V\sum_m \delta (t-mT)\,\,, H_0(\{{\cal L}_i\}) = \sum_{i=1}^d
\frac{\tau_i}{2} ({\cal L}_i+\gamma_i)^2\, \quad \quad \quad \quad \quad \quad
\quad \quad \quad \quad\quad \quad \quad \quad \\
V(\{\theta_i\}) = \cos(\theta_1) \cos(\theta_2) \cos({\alpha}) +
{1\over2} \sin(2\theta_1) \cos(2\theta_2) \sin({\alpha}) \quad \quad
\quad \quad \quad \quad\quad\quad \quad \quad \quad \quad \quad \nonumber
\end{eqnarray}
where ${\cal L}_i$ denotes the angular momentum and $\theta_i$ the conjugate angle
of one rotor. The kick period is $T$, $k$ is the kicking strength, while $\tau_i$ is
a constant inversely proportional to the moment of inertia of the rotor. The standard
KR corresponds to $d=1$ (with $\theta_2=0$) whereas for $d=2$ we have a two-dimensional 
generalization. The parameter $\alpha$ breaks TRS \cite{I90}, the parameters $\gamma_i$ 
are irrational numbers whose meaning will be explained below. The Hamiltonian (\ref{ham}) 
describes a system which is kicked periodically in time and is integrable in the 
absence of the kicking potential. The motion generated by (\ref{ham}) is classically 
chaotic and for a sufficiently strong kicking strength $k$ there is diffusion in 
momentum space with diffusion coefficient $D\equiv lim_{t\rightarrow \infty} <{\bf 
{\cal L}}^2(t)>/ t \simeq k^2/2T$ (within the random phase approximation) \cite{I90,DF88}.

If the ${\cal L}_i$ are taken mod$(2\pi m_i/T\tau_i)$ where $m_i$ are integers, 
Eq.~(\ref{ham}) defines a dynamical system on a torus. The quantum mechanics of 
this system is described by a finite-dimensional time evolution operator for one
period
\begin{equation}
\label{Uop}
U=\exp(-iH_0(\{{\cal L}_i\})T)\quad \exp(-iV(\{\theta_i\}))
\end{equation}
where we put $\hbar=1$. Upon quantization, additional symmetries associated with the 
discreteness of the momentum show up, which can be destroyed by introducing irrational 
values for the parameters $\gamma_i$'s. The most striking consequence of quantization 
is the suppression of classical diffusion in momentum space due to quantum dynamical 
localization \cite{I90,DF88}. We introduce the eigenstate components 
${\bf \Psi}_k(n)$ of the Floquet operator in the momentum representation by
\begin{equation}
\label{eigen}
\sum_m U_{mn} {\bf \Psi}_k(n) = e^{i\omega_k T} {\bf \Psi}_k(n) \quad .
\end{equation}
The quantities $\omega_k$ are known as quasi-energies, and their density is $\rho =
T/2\pi$. The corresponding mean quasi-energy spacing is $\Delta = 1/(\rho L^d)$, 
where $L$ is the linear size of the system. The Heisenberg time is $t_H=2\pi/\Delta$
while $t_D=L^2/D$ is the diffusion time (Thouless time). Now one can formally define 
a dimensionless conductance as $g=t_H/t_D = D_{k} L^{d-2}$ where $D_k=T D$ is the 
diffusion coefficient measured in the number of kicks. Four length scales are important 
here: the wavelength $\lambda$, the mean free path $l_M$, the linear extent of the 
system $L$, and the localization length $\xi$. According to Refs. 
\cite{MK95,FM94,FE95,M97,M00,SA97} the field theoretical predictions are derived 
under the conditions
\begin{equation}
\label{ftpr}
\lambda \ll l_M \ll L \ll \xi.
\end{equation}
The first condition ensures that transport between scattering events may be treated 
semiclassically. This limit can be achieved for our system (\ref{ham}) when $k
\rightarrow \infty$, $T\rightarrow 0$ while the classical parameter $K=kT$ remains 
constant. When $l_M \ll L $ as long as the motion is not localized (i.e. $ L \ll 
\xi$) it is diffusive, since a particle scatters many times before it can traverse 
the system. The resulting mean free path for our system (\ref{ham}) is $l_M\simeq 
{\sqrt D_k}$ while the localization length for $d=1$ is $\xi \simeq D_k/2$ \cite{I90} 
and for $d=2$ is $\xi \simeq l_M e^{D_k/2}$ \cite{DF88,LR57}.

Here we calculate the distribution function ${\cal P}(t=L^d |{\bf \Psi}_k(n)|^2)$ 
\cite{note1} by using a direct diagonalization of the Floquet operator (\ref{Uop}). 
The TRS is broken entirely for $\alpha=5.749$. In order to test the issue of dynamical 
correlations, we randomize the phases of the kinetic term of the evolution operator 
(\ref{Uop}) and calculate the resulting ${\cal P}(t)$. This model will be referred 
to as Random Phase KR (RPKR). Since all our eigenfunctions have the same statistical 
properties (in contrast to the Anderson cases where one should pick up only 
eigenfunctions having eigenenergies within a small energy interval \cite{UMRS00,N01}) 
we make use of all of them in our statistical analysis. The classical parameter $K$ 
is large enough in all cases to exclude the existence of any stability islands in 
phase space. The classical diffusion coefficient $D_k$ is calculated numerically by 
iterating the classical map obtained from (\ref{ham}). Below we present our numerical 
results and compare them to the predictions of Refs.~\cite{MK95,FM94,FE95,M97,M00,SA97}.

{\it 1d Kicked Rotor}. It was shown in \cite{AZ96}, that the effective field 
theory describing the semiclassical physics of the system is precisely the NLSM 
for quasi-one dimensional ($1d$) metallic wires. Such a mapping however, requires 
an averaging over an ensemble of rotors having the same classical limit. We point 
out again that in the calculations below we do not adopt such an averaging procedure.

\begin{figure}
\hspace*{-1cm}\epsfig{figure=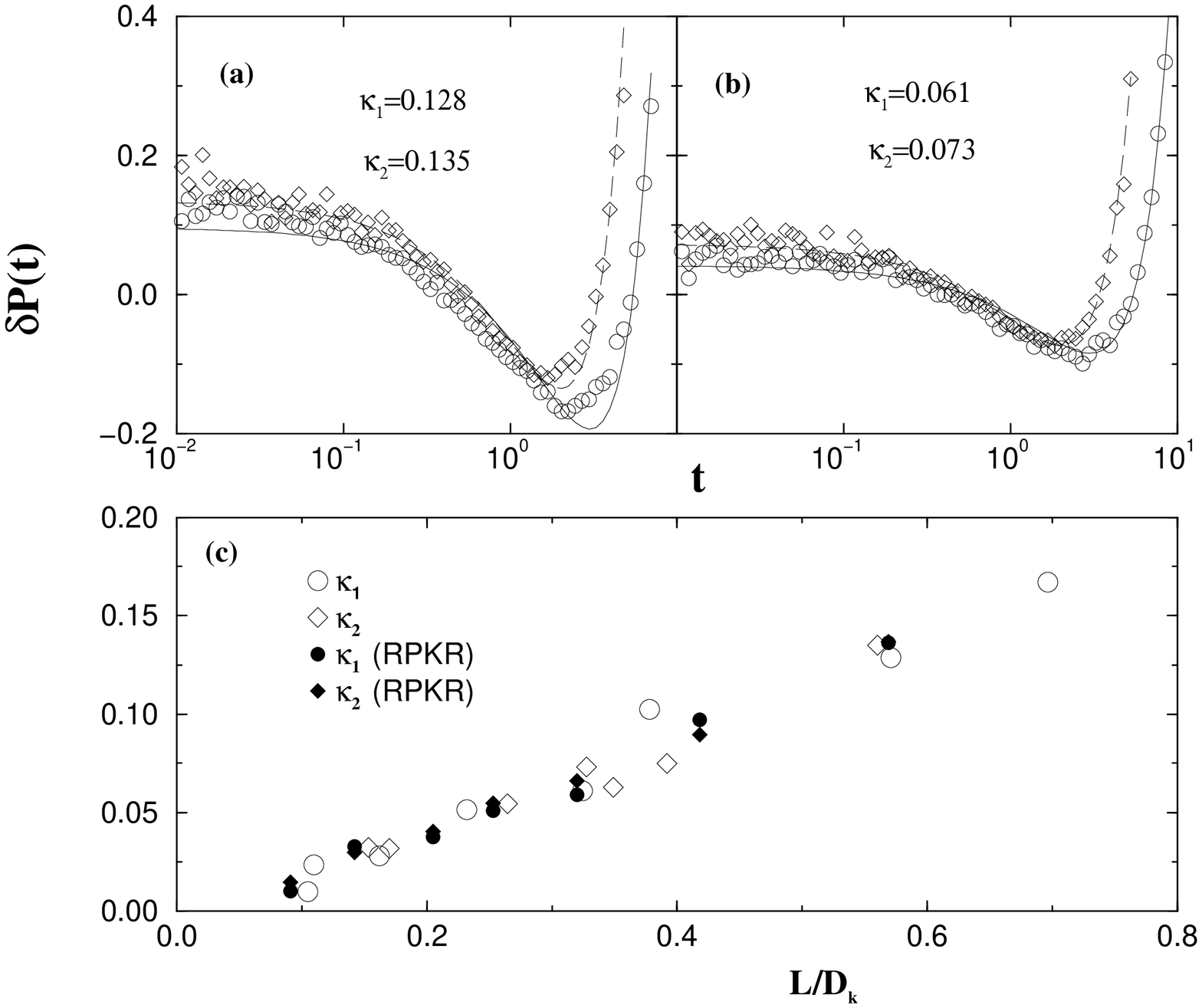,height=9cm,width=5cm,angle=270}
\noindent
\\
{\footnotesize \\
{\bf FIG. 1.} Corrections to the distribution intensities $\delta {\cal P}_{\beta}
(t)$ for the kicked rotator model i.e. Eq.~(\ref{ham}) for $d=1$. The system size 
is $L=1024$, ($\circ$) $\beta=1$, ($\Diamond$) $\beta=2$. The solid (dashed) lines
are the best fit of (\ref{q1dor}) for $\beta = 1 (2)$ to the numerical data: (a)
$D_k\approx 1800$ and (b) $D_k\approx 3150$ ; (c) Shows the extracted 
diffusion propagator $\kappa_{\beta}$ vs. $L/D_k$. }
\end{figure}

The NLSM for quasi-1d systems can be solved exactly for the distribution function
${\cal P}_\beta(t)$, using a transfer matrix approach \cite{FM94,M97,M00}. In the
ballistic regime (where $g \rightarrow \infty$) RMT is applicable and one finds ${
\cal P}_{(\beta=1)}^{RMT}(t) = \exp(-t/2)/\sqrt{2\pi t}$ and ${\cal P}_{(\beta=2)
}^{RMT}(t) = \exp(-t)$ \cite{M91}. Here $\beta$ denotes the corresponding Dyson
ensemble ($\beta=1(2)$ for preserved (broken) TRS). As localization increases, the 
deviations from the RMT results of the body and the tails of the distribution 
${\cal P }_{\beta}(t)$ become noticeable and can be parameterized by a single 
parameter which is the dimensionless conductance $g=D_k/L$.

For $t< \sqrt{D_k/L}$, according to all studies \cite{FM94,FE95,M97} ${\cal P} (t)$
is just the RMT result with polynomial corrections in powers of $L/D_k$, i.e. $
{\cal P}_{\beta}(t)= {\cal P}_{\beta}^{RMT}(t)[1+ \delta {\cal P}_\beta(t)]$. The 
leading term of this expansion is given by
\begin{eqnarray}
\label{q1dor}
\delta {\cal P}_{\beta}(t)  &\simeq & \kappa
\left\{
\begin{array}{ll}
3/4-3t/2+t^2/4 &\mbox{for$\,\beta=1$}\,\\
1-2 t+t^2/2    &\mbox{for$\,\beta=2$}\,
\end{array}
\right \},
\end{eqnarray}
where $\kappa\sim 1/g$ is the $1d$ diffusion propagator, which is identical for
$\beta=1$ and $\beta=2$ since it is a classical quantity.

In Fig.~1a,b we report our numerical results for $\delta {\cal P}_{\beta}(t)$ for
two representative values of $D_k$. One can clearly see that the agreement with
the theoretical prediction (\ref{q1dor}) becomes better as $D_k$ increases. This
is due to the fact that by increasing $D_k$ we are approaching the semiclassical
region and therefore Eqs.~(\ref{ftpr}) are better satisfied. At the same time 
higher order corrections in $\delta {\cal P}_{\beta}(t)$ become negligible with
respect to the leading term given by Eq.~(\ref{q1dor}). The resulting $\kappa_1$
and $\kappa_2$ obtained by the best fit of our data to Eq.~(\ref{q1dor}) are
found to be equal and in excellent agreement with the theoretical value (see 
Fig.~1c). We therefore conclude, that in a generic dynamical system, the only 
parameter that controls the shape of the deviations $\delta {\cal P}_{\beta}(t)$ 
is the classical diffusion propagator. Moreover, our results are in excellent 
agreement with the recent NLSM predictions derived in the framework of diffusive 
disordered systems. Finally in Fig.~1c we also report the outcome of the RPKR 
model. The results remain essentially the same indicating that ${\cal P}_{\beta}
(t)$ for quasi-$1d$ systems are insensitive to dynamical correlations.

\begin{figure}
\hspace*{-1cm}\epsfig{figure=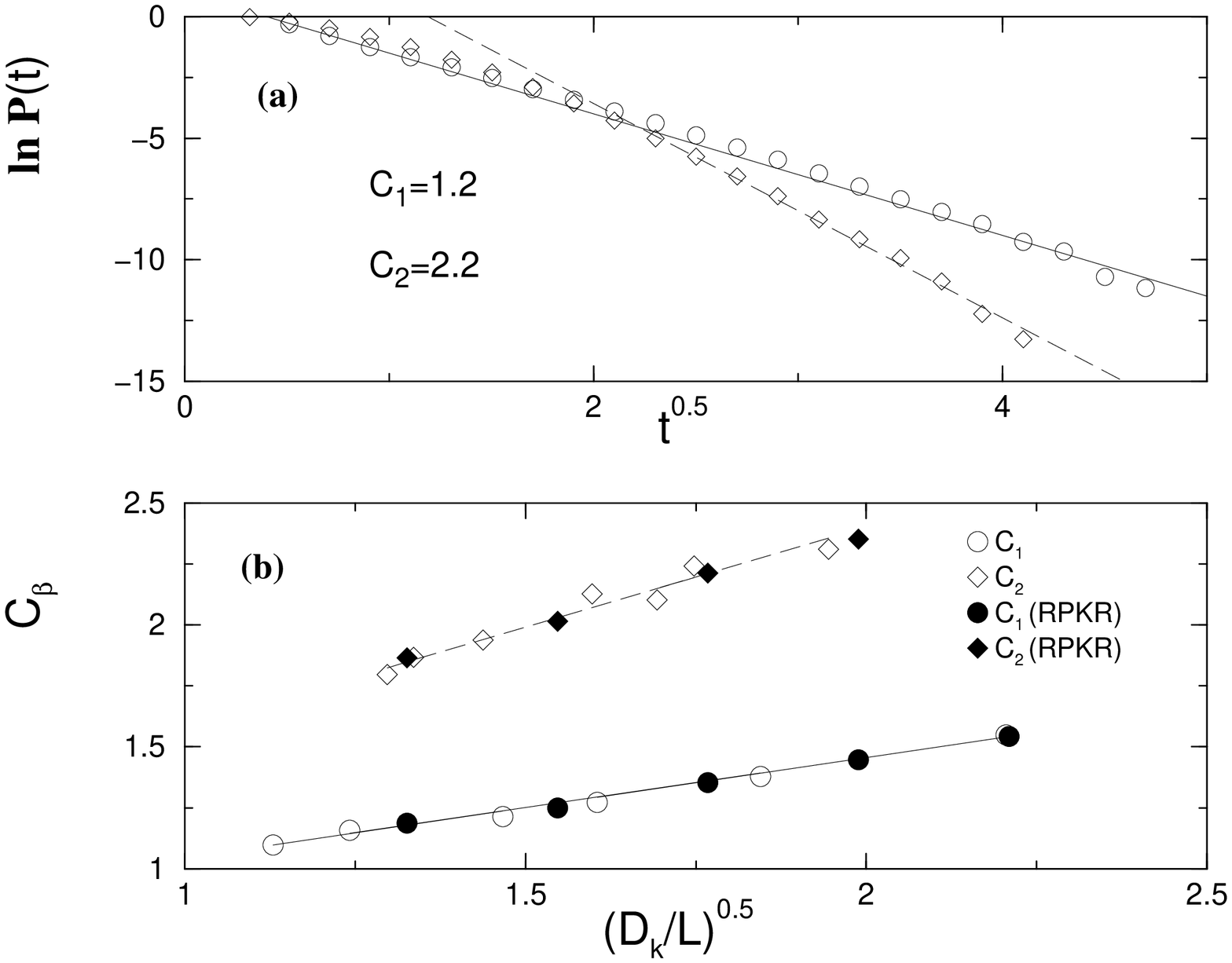,height=9cm,width=5cm,angle=270}
\noindent
\\
{\footnotesize \\
{\bf FIG. 2.}
(a) Tails of the distribution ${\cal P}_{\beta} (t>D_k/L)$ for the model (\ref{ham}) 
for $d=1$ with $L=1024$,
$D_k\simeq 2625$ and for $\beta=1$ ($\circ$) and $\beta=2$ ($\Diamond$). The solid 
(dashed) lines are the best fit of (\ref{q1dt}) for $\beta=1(2)$ to our data; (b) 
Coefficients $C_{\beta}$ vs. $\sqrt{D_k/L}$. The solid (dashed) lines are the 
best fits to $C_{\beta}= A_{\beta} \sqrt{D_k/L}+ B_{\beta}$ for $\beta=1 (2)$.}
\end{figure}

The tail of the distribution ($t> D_k/L$) deviates strongly from the RMT
prediction
and has a stretched exponential form \cite{FM94,FE95,M97}
\begin{equation}
\label{q1dt}
{\cal P}_\beta(t) \simeq A_\beta \exp(-2 C_{\beta}\sqrt{t}) ,\quad
C_{\beta}=\beta\sqrt{D_k/L}
\end{equation}
where $A_{\beta}$ is a symmetry dependent constant.
Our numerical results agree nicely with Eq.~(\ref{q1dt}). In Fig.~2a we present an
example of ${\cal P}_\beta(t)$. By fitting our data to Eq.~(\ref{q1dt}) the 
coefficients $C_1,C_2$ can be
extracted. In Fig.~2b we report the extracted stretched exponential coefficients
$C_{\beta}$ from the best fit of (\ref{q1dt}) as a function of the square root of the
dimensionless conductance $g=D_k/L$. A nice linear behavior is observed. The best linear
fit $C_{\beta}= A_{\beta}{\sqrt{ D_k/L}}+B_{\beta}$ yields, $A_{\beta=1}=0.41\pm 0.05$
and $A_{\beta=2}= 0.82\pm 0.05$. The resulting ratio $R=A_2/A_1 =2$ is in excellent
agreement with the theoretical prediction (\ref{q1dt}). We have also calculated the
stretched exponential coefficients $C_{\beta}$ for the RPKR model. The results for
various $D_k$ values are summarized in Fig.~2b and show a nice agreement with the 
results obtained from the real Hamiltonian.

{\em $2d$ Kicked Rotor}. According to Ref.~\cite{FM94}, corrections to the 
body of ${\cal P}_{\beta}^{RMT}$ are still given by Eq.~(\ref{q1dor}), but now $\kappa$
is the $2d$ diffusion propagator.

Figures~3a,b show corrections to ${\cal P}_{\beta}^{RMT}$ for $g=D_k\gg 1$ for two
representative values of $D_k$. We find again that the form of the deviations are
very well described by Eq.~(\ref{q1dor}) and the agreement becomes better for larger
values of the diffusion constant. In Fig.~3c we summarize our results for various
$D_k$ values. The extracted $\kappa_{\beta}$ values are obtained by the best fit of 
the data to Eq.~(\ref{q1dor}). Again we find that $\kappa_{\beta}$ depends linearly 
on $1/D_k$.  However, contrary to the $1d$-KR, here $\kappa_1$ and $\kappa_2$, are 
different. Moreover the best fit with $\kappa_{\beta}=A_{\beta}D_k^{-1}+ B_{\beta}$ 
yields $A_{\beta=1}=5.44\pm0.03$ and $A_{\beta=2}=10.84\pm 0.04$ indicating that the 
ratio $R=A_2/A_1$ is close to $2$, a value that could be explained on the basis of 
ballistic effects \cite{M00,UMRS00}. Taking the latter into account leads to an 
additional term in the classical propagator $\kappa_{\beta}=\kappa_{diff}+{\beta 
\over 2} \kappa_{ball}$. The first term is the one discussed previously and is 
associated with long trajectories which are of diffusive nature while the latter
one is associated with short ballistic trajectories which are self-tracing \cite{M00}.
Thus, when $\kappa_{diff} \ll \kappa_{ball}$ we get $R=2$. The calculation with the RPKR 
model shows, however, that the corresponding ratio is $R\simeq 1$ in agreement with the 
theoretical prediction for disordered systems with a pure diffusion. This indicates 
that dynamical correlations can be important in the $2d$ case.

\begin{figure}
\hspace*{-1cm}\epsfig{figure=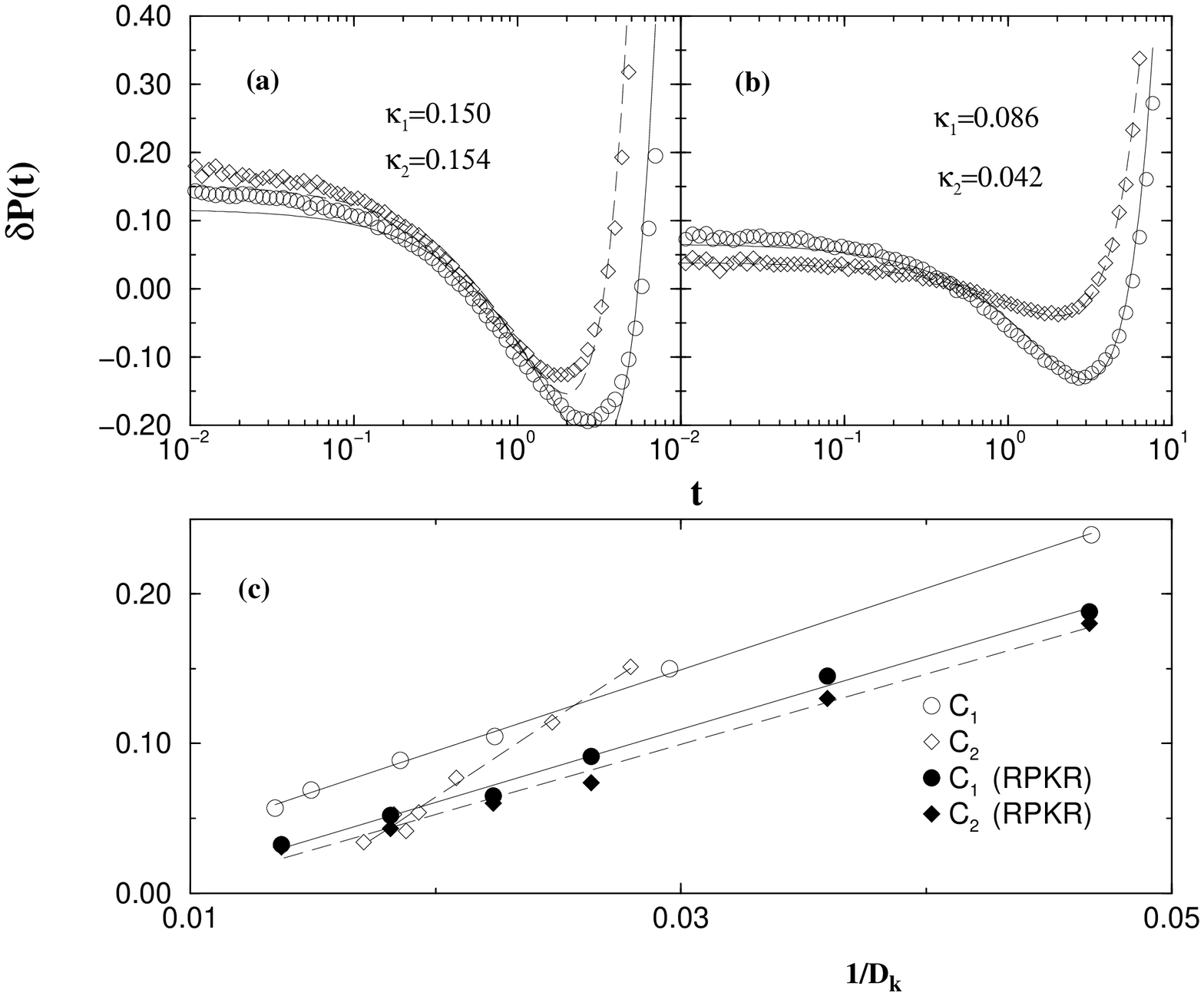,height=9cm,width=5cm,angle=270}
\noindent
\\
{\footnotesize \\
{\bf FIG. 3.}
Corrections to the distribution intensities $\delta {\cal P}_{\beta} (t)$ for the
kicked rotator model (\ref{ham}) for $d=2$. The system size is $L=90$, ($\circ$) 
$\beta=1$, ($\Diamond$) $\beta=2$. The solid (dashed) lines are the best
fit of (\ref{q1dor}) for $\beta=1 (2)$ to the numerical data: (a) $D_k\approx 34$ and
(b) $D_k\approx 53$ ; (c) Fit parameters $\kappa_{\beta}$ vs. $D_k^{-1}$. The 
solid (dashed) lines are the best fits to $\kappa_{\beta}= A_{\beta} D_k^{-1}+ 
B_{\beta}$ for $\beta=1 (2)$.}
\end{figure}

For the tails of the distributions, the result of the NLSM within a saddle-point
approximation \cite{FE95,M00} is
\begin{equation}
\label{nlsmt}
{\cal P}_\beta(t) \simeq \exp[-C_{\beta}^{\sigma } (\mbox{ln} t)^2 ] , \quad
C_{\beta}^{\sigma} = {\beta\pi^2\rho\over2} {D\over\mbox{ln}(L/ l)}.
\end{equation}
Note that the decay in the  tails of Eq.~(\ref{nlsmt}) depends on $\beta$, as in
the $1d$-KR case (see Eq.~(\ref{q1dt})). Recently, a DOF method was used to calculate
the tails of ${\cal P}_{\beta}(t)$ \cite{SA97}. It was found that the tails are 
still given by Eq.~(\ref {nlsmt}) but with a log-normal coefficient $C$ which is
independent of the parameter $\beta$ :
\begin{equation}
\label{ofm}
C^{\rm DOF} = \pi^2 \rho {D\over\mbox{ln}(L/\lambda)}.
\end{equation}

\begin{figure}
\hspace*{-1cm}\epsfig{figure=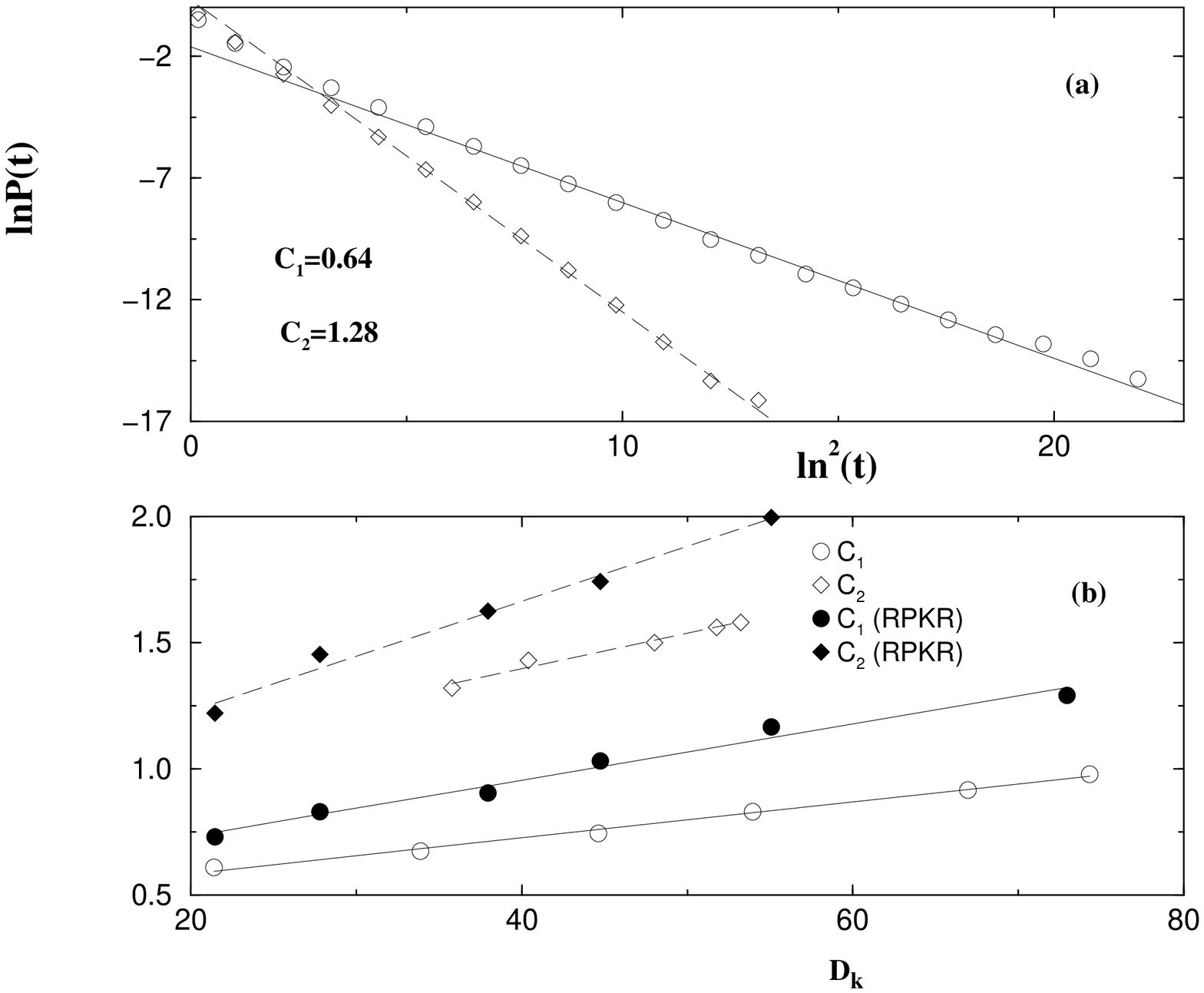,height=9cm,width=5cm,angle=270}
\noindent
\\
{\footnotesize \\
{\bf FIG. 4.}
(a) Tails of the distribution ${\cal P}_{\beta=1}(t>D_k)$ for the model (\ref{ham}) 
for $d=2$ and $D_k\simeq
35$. The system size is $L=80$, ($\circ$) $\beta=1$, ($\Diamond$) $\beta=2$.
The solid (dashed) lines are the best fit of (\ref{nlsmt}) for $\beta=1 (2)$ to 
the numerical data; (b) Fitted log-normal coefficients $C_{\beta}$ versus the 
classical diffusion coefficient $D_k$. The solid (dashed) lines are the best fits 
to $C_{\beta}= A_{\beta} D_k+ B_{\beta}$ for $\beta=1 (2)$. }
\end{figure}

Figure~4a shows a representative case of ${\cal P}_{\beta=1}(t> D_k)$. The tails
show a log-normal behavior predicted by Eq.~(\ref{nlsmt}). In Fig.~4b we report
the log-normal coefficients $C_{\beta}$ extracted from the best fit to our numerical 
data, versus the classical diffusion coefficient. A pronounced linear behavior 
is observed in agreement with both theories. However one clearly sees that $C_1$ 
differs from $C_2$ in contrast to the DOF prediction (\ref{ofm}) and to recent numerical
calculations done for the $2d$ Anderson model \cite{UMRS00}. We point out here that in
\cite{UMRS00} the authors were not able to go to large enough values of conductance 
$g$ (in comparison to our study) where the theory can really be tested. In contrast, 
the NLSM predicts a value of $2$ for the ratio $R=C_2^{\sigma}/C_1^{\sigma}$. We note 
that $C_{\beta}^{\sigma}$ is only the leading term in $D_k$. In order to calculate
this ratio, we performed a fit to our data with $C_{\beta}=A_{\beta} D_k+ B_{\beta}$.
The resulting ratio was found to be $R=A_2/ A_1= 1.97 \pm 0.03$ in perfect agreement
with the NLSM predictions. Finally in Fig.~4b we also present our results for the 
RPKR model (using the same data as the one in Fig.~3d). Again we found that the 
ratio $R= 1.96\pm 0.03 \approx 2$. Thus ${\cal P}(t> D_k)$ depends on TRS and is 
described by the NLSM.

In summary, we have performed a detailed numerical analysis of the statistical 
properties of the wavefunction intensities ${\cal P}(t)$ of the standard KR on a 
torus and its $2d$ generalization. Based on these results, we concluded that the 
distribution ${\cal P}(t)$ of generic quantum {\it dynamical} systems with diffusive 
classical limit is affected by the existence of {\it prelocalized} states. The 
deviations from RMT are well described by field theoretical methods developed for 
disordered systems. In particular, in a clarifying way we have resolved the 
controversy between DOF and NLSM by demonstrating that the dependence of the tails 
of ${\cal P}_{\beta}(t)$ on TRS is described correctly by the latter theoretical 
approach.

We acknowledge useful discussions with L. Kaplan.
%=======================================================================================

\begin{figure}
\end{figure}

\end{multicols}
\end{document}